\documentclass[11 pt, onecolumn]{article}
\usepackage[margin=1.1in]{geometry}

\usepackage{amsmath}
\usepackage{bm}	
\usepackage{graphicx}

\newtheorem{Proposition}{Proposition}

\usepackage{graphicx}
\usepackage{multirow}
\usepackage{mdwlist}

\usepackage{threeparttable}

\usepackage{amsfonts}
\usepackage{amsmath}
\usepackage{hyperref}
\usepackage{bbm}
\usepackage{float}
\usepackage{caption}
\usepackage{subcaption}
\usepackage{xcolor}
\usepackage{mathtools}
\usepackage{tikz}

\usetikzlibrary{shapes.geometric,arrows}
\usepackage{enumerate}
\usepackage{multicol}
\usepackage{stackrel}

\usepackage{subcaption}
\usepackage{amssymb}
\usepackage[utf8]{inputenc}
\usepackage{booktabs,lipsum}
\usepackage{hyperref}
\usepackage{graphicx}
\usepackage{epstopdf}

\begin{document}
\title{\LARGE \bf Analysis of a Stochastic Model for Coordinated Platooning of Heavy-duty Vehicles}

\author{Xi Xiong, Erdong Xiao, and Li Jin
\thanks{This work was supported in part by NYU Tandon School of Engineering and C2SMART Department of Transportation Center. The authors appreciate the discussion with Profs. Pravin Varaiya, Saurabh Amin, Karl H. Johansson, and Zhong-Ping Jiang.}
\thanks{X. Xiong and L. Jin are with the Department of Civil and Urban Engineering, and E. Xiao is with the Department of Mechanical and Aerospace Engineering, New York University Tandon School of Engineering, Brooklyn, NY, USA, emails: xi.xiong@nyu.edu, lijin@nyu.edu, ex276@nyu.edu.}%
}
\newcommand*{\QEDA}{\hfill\ensuremath{\blacksquare}}%

\maketitle

\begin{abstract}
Platooning of heavy-duty vehicles (HDVs) is a key component of smart and connected highways and is expected to bring remarkable fuel savings and emission reduction.
In this paper, we study the coordination of HDV platooning on a highway section.
We model the arrival of HDVs as a Poisson process.
Multiple HDVs are merged into one platoon if their headways are below a given threshold.
The merging is done by accelerating the following vehicles to catch up with the leading ones.
We characterize the following random variables: (i) platoon size, (ii) headway between platoons, and (iii) travel time increment due to platoon formation.
We formulate and solve an optimization problem to determine the headway threshold for platooning that leads to minimal cost (time plus fuel).
We also compare our results with that from Simulation of Urban MObility (SUMO).
\end{abstract}


\section{Introduction}

Platooning of heavy-duty vehicles (HDVs) has the potential for remarkable efficiency and environmental benefits~\cite{hor+var00,bess+16procieee,litman2017autonomous}.
A typical platooning procedure on a highway network involves two steps, viz. (i) coordinating the time when multiple HDVs arrive at a junction and (ii) merging the HDVs into a platoon at the junction \cite{varaiya1993smart,larson2015distributed}.
Although much progress has been made recently on vehicle-level control algorithms to achieve safe and efficient merging \cite{coogan15interconnected,besselink2017string,gao2017data,hu2018plug}, we still lack a realistic and tractable framework to evaluate and design large-scale coordinating strategies.

Previous work on platoon formation typically focused on development of efficient scheduling algorithms that coordinate the movement of HDVs to facilitate platooning.
Larson et al. \cite{larson2015distributed} proposed a network of distributed platooning coordinators that manage formation of platoons at highway junctions.
Van de Hoef et al. \cite{van2018fuel} considered the en-route formation of platoons.
However, the previous work mainly focused on analysis at the vehicle level and in deterministic settings.
However, to support large-scale platooning operations, a system-level model that captures the randomness in the HDVs' movement is needed.
Such a model enables us to better understand the impact of various platooning strategies on properties of the consequent HDV platoons and to design platooning strategies with performance guarantees.

\begin{figure}
  \centering
  \includegraphics[width=0.47\textwidth]{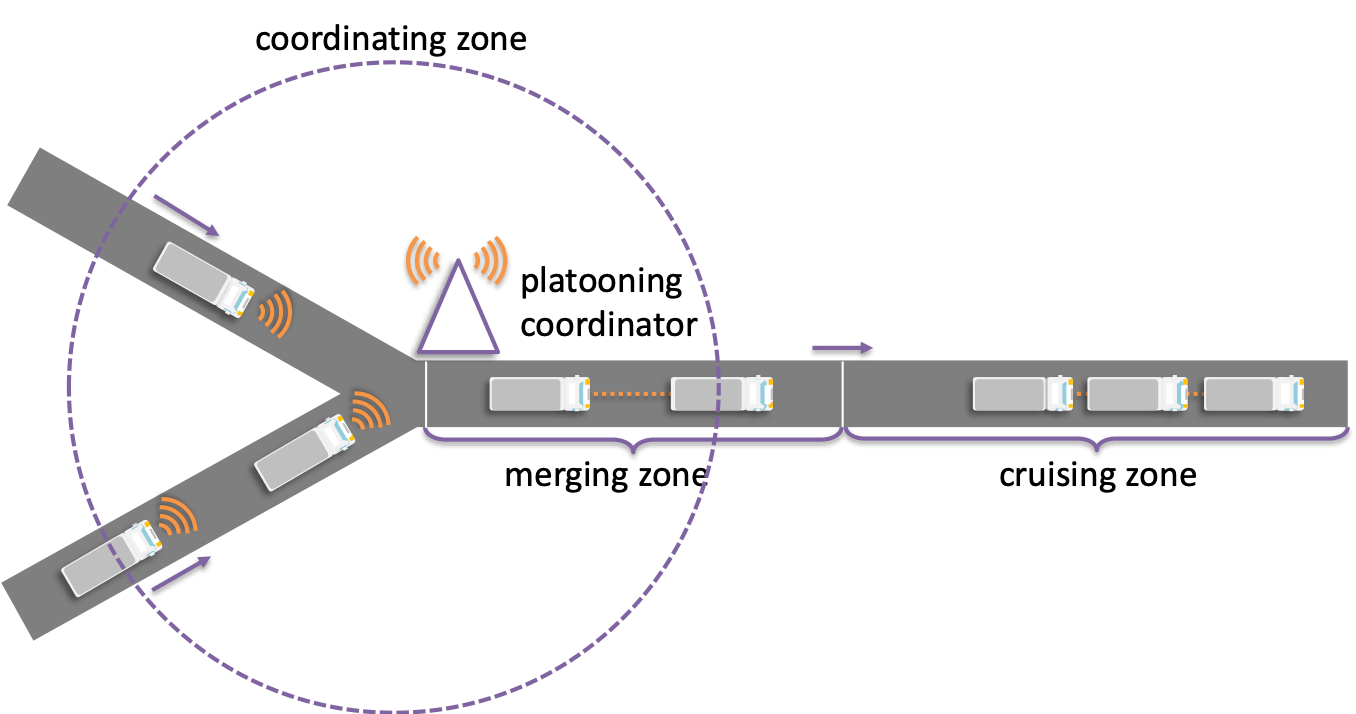}
  \caption{Coordinated platooning on a highway section.}\label{fig:color_figure}
\end{figure}

In this paper, we develop an analytical model for the coordinated platooning process on a highway section and optimize the platooning strategy using this model.
Fig.~\ref{fig:color_figure} illustrates the setting that we consider.
A platooning coordinator is situated at the entrance to a highway (e.g. a junction).
The coordinator is able to collect movement information of every HDV within a coordinating zone, the size of which is significantly larger than the original spacings between HDVs before getting platooned.
HDVs randomly enter the coordinating zone and arrive at the entrance as a homogeneous Poisson process.
The entrance is followed by a pre-specified merging zone, where HDVs form platoons.
A coordinator at the entrance instructs multiple HDVs to platoon if their headways are below a certain threshold, which is the major decision variable considered in this paper.
To complete the platooning, the coordinator instructs the involved vehicles to coordinate their speeds within the merging zone so that they can form a platoon.
HDVs leave the merging zone and travel on the rest of the highway (cruising zone) as platoons.
The above model can be viewed as a link-level refinement and a stochastic extension of the model considered in \cite{larson2015distributed}.

To obtain system-level insights, we develop a stochastic model that is closely related to random geometric graphs and Poisson point processes \cite{penrose2003random}.
Such a relation is very natural, since HDVs on a highway essentially form a one-dimensional spatial point process, and platooning is essentially connecting and clustering nearby HDVs.
Researchers have used similar models to study the connectivity of the communications network (vehicular ad hoc network) over connected vehicles with random headways \cite{wu2004analytical,yin2013vehicle}.
A key distinction between our model and the classical Poisson point process is that we also consider relocation of points (HDVs) to form connected components (platoons), while classical models only connect but do not relocate points.
Note that our model is also related to but distinct from queuing systems (even those with batch services) in that our model, although it involves a Poisson arrival process, does not have a service process or incorporate the influence of background vehicles, and thus does not involve waiting due to limited service capacity.

We use the stochastic model to characterize platoon sizes, headways between formed platoons, and time increment due to platooning.
We are able to analytically compute the distribution of platoon size and the expected values of between-platoon headways and time increment (Propositions~\ref{prp_size}--\ref{prp_time}).
Our results imply that platoon sizes, between-platoon headways, and time increments all increase with the headway threshold for platooning.
Our results are also quantitatively consistent with the empirical prediction from the Simulation of Urban MObility (SUMO), a widely used tool for traffic simulation \cite{krajzewicz2002sumo}.

Characterization of the platooning process helps us understand the relation between the headway threshold for platooning and the system-wide cost, which is the sum of three parts: (i) traverse time in the merging zone, (ii) fuel consumption in the merging zone, and (iii) fuel consumption over the cruising zone, i.e. the highway section beyond the merging zone.
By analyzing our stochastic model, we are able to analytically characterize these three costs by studying the behavior of between-platoon headways, platoon sizes, and increment of between-vehicle headways induced by platooning.
These quantities are also relevant for studying the interaction between HDV platoons and background traffic sharing common highway segments \cite{jin2018modeling}.

Based on these results, we formulate a simple optimization problem for determining the optimal headway threshold for platooning under a variety of scenarios.
In general, the coordinator may specify the platooning process flexibly: leading vehicles can be decelerated, following vehicles can be accelerated, or a mixture of acceleration/deceleration can be applied.
In this paper, we follow \cite{larson2015distributed} and only consider acceleration; i.e. platoons are formed by following vehicles' catching up. Nevertheless, our results can be readily extended to the deceleration or the mixed cases.
By solving the optimization problem, we find that the optimal platooning strategy parametrized by the headway threshold strongly depends on the length of the highway section: if the highway is long enough, then higher threshold and longer platoons are favorable; otherwise, short platoons or even no platooning is favorable.
Beyond this intuitive insight, our model can quantitatively support online coordinated platooning.


The rest of this paper is organized as follows.
In Section~\ref{sec_model}, we introduce the stochastic model.
In Section~\ref{sec_analysis}, we derive key properties of the platooning process and compare our results to empirical results from the SUMO.
In Section~\ref{sec_optimize}, we formulate and solve an optimization problem for coordinated platooning.
In Section~\ref{sec_conclude}, we summarize the conclusions and propose several directions for future work. 
\section{Modeling platooning process}
\label{sec_model}

\begin{figure}[ht]
  \centering
  \includegraphics[width=0.45\textwidth]{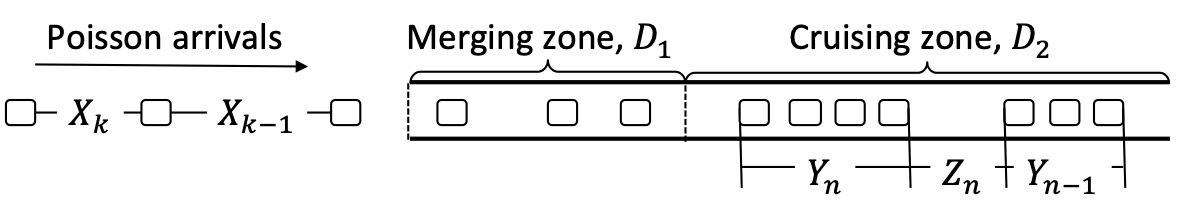}
  \caption{Between-vehicle headway $X_k$, platoon size $Y_n$, and between-platoon headway $Z_n$.}\label{fig:merge_zone}
\end{figure}

Consider a highway segment subject to upstream HDV arrivals.
Without loss of generality, we assume zero initial condition.
Let $S_0=0$ and $S_k$ be the time at which the $k$th vehicle enters the highway. The inter-arrival times $X_k=S_k-S_{k-1}$, $k=1,2,\ldots$ are independent and identically distributed random variables (rv s), which follow the exponential distribution with parameter $\lambda$.
Practically, $\lambda$ is the average arrival rate of vehicles.
Platooning is completed over a merging zone following the entrance to the highway, and the vehicles travel in platoons over the rest part of the highway; see Fig.~\ref{fig:merge_zone}.

Let $r$ be the headway threshold for platooning. That is, if $X_k\le r$, then the $(k-1)$th and the $k$th vehicles merge into a platoon. If the $(k-1)$th vehicle is already in a platoon, then the $k$th vehicle merges into that platoon. Since we focus on the system-level control of platooning, the impacts of velocity and distance during merging can be ignored. After the merging process, we assume that the headways between vehicles within a platoon are very small and approximate as 0 to simplify calculation; this assumption is valid if the within-platoon headway is significantly smaller than the original headways (i.e. $X_k$). When a following vehicle merges with a leading vehicle or a leading platoon, there are multiple ways to coordinate their movements:
(i) acceleration -- the following vehicle accelerates while the leading vehicle/platoon maintains its original speed;
(ii) deceleration -- the leading vehicle/platoon decelerates while the following vehicle maintains its original speed;
(iii) cooperation -- the leading vehicle/platoon decelerates while the following vehicle accelerates.
In this article, we only consider case (i). Nevertheless, our approach applies to cases (ii) and (iii) as well.

\begin{figure}[b!]
  \centering
  \includegraphics[width=0.25\textwidth]{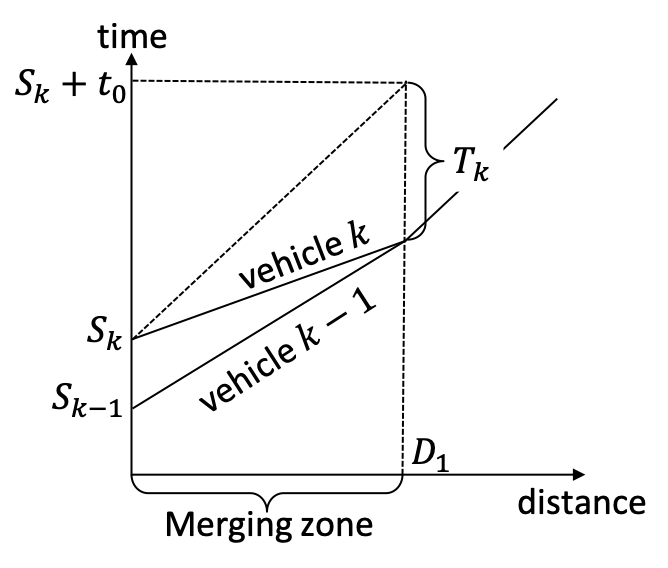}
  \caption{Time increment $T_k$ due to acceleration for platooning.}\label{fig:Tk}
\end{figure}

We use $Y_n\in\{1,2,\ldots\}$ to denote the number of vehicles in the $n$th platoon, and $Z_n$ to denote the headway between the $(n-1)$th and the $n$th platoons.
The time for the $k$th vehicle to traverse the merging zone consists of two parts, viz. the nominal traverse time $t_0$, which is identical for all vehicles, and the increment $-T_k\le0$ due to acceleration for merging.
Hence, the time at which the $k$th vehicle leaves the merging zone is $S_k+t_0-T_k$; see Fig.~\ref{fig:Tk}.
Our subsequent analysis will focus on $Y_n$, $Z_n$, and $T_k$, which are three important quantities that characterize the platoons formed in the merging zone.
Throughout this article, we assume that the process is in steady state. That is, the between-platoon headyway $Z_n$ is a rv taking value from $[0,\infty)$.

Furthermore, note that the actual headway between the $k$th and the $(k-1)$th vehicles when the $k$th vehicle enters the highway may be different from $X_k$ due to acceleration of the $(k-1)$th vehicle.
We assume that the decision on whether to merge only depends on the original headway $X_k$, regardless of the actual headway when the $k$th vehicle enters the highway.
This assumption is valid if the radius of the communication zone is significantly larger than $\mathrm E[X_k]$; in this case, the coordinator has an almost global knowledge of the movement of every vehicle that would be involved.

\section{Analyzing platoon characteristics}
\label{sec_analysis}

When operating platoons on the highway, we need to identify some essential characteristics, including platoon size, headway between platoons and time increment of platooning. In this section, we present related propositions under the assumed arrival and platooning processes. Note that all the results consider the processes in the steady state. We also compare the model-predicted results with empirical results obtained from a micro-simulation model.

\subsection{Analysis based on stochastic model}
First, we derive the distribution of the number of vehicles in a platoon:
\begin{Proposition}[Platoon size]
\label{prp_size}
In the steady state, the number of vehicles in a platoon $Y$ follows the probability mass function (PMF):
\begin{align*}
\ p_Y(y)= e^{-\lambda r}(1-e^{-\lambda r})^{y-1} \quad y=1,2,3, \cdots
\end{align*}
\end{Proposition} \

\begin{figure}[ht]
  \centering
  \includegraphics[width=0.47\textwidth]{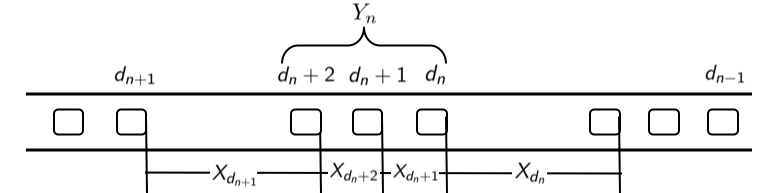}
  \caption{Illustration of platoons}\label{fig:platoon_figure}
\end{figure}

\emph{Proof}
We assume the index of the first vehicle in the $n$th platoon is $D_n$, and the platoon size is $Y_n$. Since $D_n$ is independent of $X_{D_n}$, we replace $D_n$ with $d_n$ to simplify notations. Then the index list in platoon $n$ is $d_n, d_n+1, d_n+2, \cdots, d_n+(Y_n-1)$. The vehicle index list is shown in Fig. \ref{fig:platoon_figure}. We can make sure that the inter-arrival time $X_{d_{n+1}} > r$, which enables platoon $n$ and platoon $n+1$ to be two different clusters of vehicles. The probability of platoon size being $y$ is 
\begin{align*}
 \Pr \left\{Y_n = y \right\}  & = \Pr \left\{ X_{d_{n+1}}>r, X_{d_n+1} \le\ r, X_{d_n+2} \le\ r, \right. \\
 									  & \quad \left. \cdots, X_{d_n+(y-1)} \le\ r \right\} \\
									  & = \Pr \left\{X_{d_{n+1}}>r \right\} \Pr \left\{X_{d_n+1} \le\ r \right\} \\
									  & \quad  \Pr \left\{X_{d_n+2} \le\ r \right\} \cdots \Pr \left\{X_{d_n+(y-1)} \le\ r \right\}.
\end{align*}

If we assume the probability of merging is $p$, we can show that:
\begin{align*}
\Pr \left\{Y_n = y \right\} = (1-p) {p}^{y-1}.
\end{align*}
We can see that the platoon size follows a geometric distribution. 

In our case, we use Poisson process to model the vehicle arrival scenario, and the probability of merging is $1 - e^{-\lambda r}$.
Then the distribution of platoon size is shown as follows:
\begin{align*}
\Pr \left\{Y_n = y \right\} = e^{-\lambda r}(1-e^{-\lambda r})^{y-1}.
\end{align*}

\hfill $\blacksquare$

Next, we compute the expected headway between two platoons:

\begin{Proposition}[Headway between platoons]
The expectation of headway $Z$ between platoons is:
\begin{align*}
\mathrm{E}[Z] = \frac{e^{\lambda r}}{\lambda}.
\end{align*}
\end{Proposition} \

\emph{Proof}
The headway changes in acceleration and deceleration cases are symmetrical. We then use the acceleration process to model the headway distribution. In the acceleration case, the following vehicle accelerates while the leading vehicle  maintains its original speed. Let $Z_n$ be the headway between the first vehicle in platoon $n$ and the last vehicle in platoon $n-1$. $Z_n$ increases with the number of vehicles in platoon $n-1$. The cumulative distribution function (CDF) of $Z_n$ is
\begin{align*}
\ F_{Z_n}(z) 
	   & =\sum_{y=1}^{\infty} \Pr \left\{Z_n \leq z | Y_{n-1}=y \right\} \Pr \left\{Y_{n-1}=y \right\}.
\end{align*}
We use the same notations in Proposition 1, then the vehicle index list in platoon $n-1$ is $d_{n-1}, d_{n-1}+1, d_{n-1}+2, \cdots, d_{n-1}+(Y_{n-1}-1)$. Since the intra-platoon headway $h_0$ is significantly smaller than the original headways (i.e. $X_{d_n}$), we can simplify calculation by assuming $h_0 \approx 0$. Then $Z_n$ can be expressed as
\begin{align*}
Z_n =  X_{d_n} + X_{d_{n-1}+1} + 
\cdots + X_{d_{n-1}+(Y_{n-1}-1)}.
\end{align*}

We can calculate the expectation of between platoon headways by the law of total expectation:
\begin{align*}
\mathrm{E}[Z_n] 
				& = \sum_{y=1}^{\infty} \Pr \left\{Y_{n-1}=y \right\} \Big( \mathrm{E}[X_{d_n} | Y_{n-1}=y] + \\
				& \quad \ \mathrm{E}[X_{d_{n-1}+1}|Y_{n-1}=y] + \cdots +  \\
				& \quad \ \mathrm{E}[X_{d_{n-1}+(y-1)}|Y_{n-1}=y] \Big).
\end{align*}

We have shown that:
\begin{align*}
\Pr \left\{Y_{n-1} = y \right\} & = \Pr \left\{ X_{d_{n}}>r, X_{d_{n-1}+1} \le\ r, \right. \\
& \quad \left.  \cdots, X_{d_{n-1}+(y-1)} \le\ r \right\}.
\end{align*}

Inter-arrival time $X_{d_n}, X_{d_{n-1}+1}, \cdots, X_{d_{n-1}+(y-1)}$ are independent and identical distributed. We can simplify the calculation of expectation by
\begin{align*}
\mathrm{E}[Z_n] & = \sum_{y=1}^{\infty} \Pr \left\{Y_{n-1}=y \right\} \Big( \mathrm{E}[X_{d_n} | X_{d_{n}}>r] +\\
				& \quad \ \mathrm{E}[X_{d_{n-1}+1} | X_{d_{n-1}+1} \le\ r] + \cdots +  \\
				& \quad \ \mathrm{E}[X_{d_{n-1}+(y-1)}|X_{d_{n-1}+(y-1)} \le\ r] \Big) \\
				& = \sum_{y=1}^{\infty} \Pr \left\{Y_{n-1}=y \right\} \Big( \mathrm{E}[X_{d_n} | X_{d_{n}}>r] +\\
				& \quad \ (y-1) \mathrm{E}[X_{d_n} | X_{d_{n}} \leq r] \Big).
\end{align*}
In our Poisson arrival scenario, we can show that:
\begin{align*}
& \mathrm{E}[X_{d_n} | X_{d_{n}}>r] = \frac{\int_{r}^{\infty} x \lambda e^{-\lambda x} dx}{e^{-\lambda r}} = \frac{1}{\lambda} + r, \\
& \mathrm{E}[X_{d_n} | X_{d_{n}} \leq r] = \frac{\int_{0}^{r} x \lambda e^{-\lambda x} dx}{1-e^{-\lambda r}} = \frac{1}{\lambda} - \frac{r}{e^{\lambda r} - 1}.
\end{align*}

Then we can derive the expectation of between platoon headway as 

\begin{align*}
\mathrm{E}[Z_n] & = \sum_{y=1}^{\infty} e^{-\lambda r}(1-e^{-\lambda r})^{y-1} \Bigg[ \left( \frac{1}{\lambda} + r\right) + \\
				& \quad \ (y-1)\left(\frac{1}{\lambda} - \frac{r}{e^{\lambda r} - 1}\right)\Bigg] \\
				& = \frac{e^{\lambda r}}{\lambda}.
\end{align*}

\hfill $\blacksquare$

Finally, we compute the expected time increment experienced by one vehicle due to platooning:
\begin{Proposition}[Time reduction of platooning]
\label{prp_time}
Also, the travel time cost of platooning in acceleration and deceleration cases are symmetrical. In the acceleration process, time increment is negative. We use positive time reduction to represent benefits of platooning. The  expectation of time reduction $T$ is
\begin{align*}
\mathrm{E}[T] = \frac{1}{\lambda} e^{\lambda r} - r - \frac{1}{\lambda}.
\end{align*}
\end{Proposition} \

\emph{Proof}
In this case, we assume the vehicle before $k$th vehicle to be $k-1$ and the vehicle after $k$th vehicle to be $k+1$. Vehicle $k-1$ belongs to platoon $n-1$. Before $k$th vehicle's arriving, the number of vehicle in platoon $n-1$ is a rv $M$ taking value from $\{1,2,\ldots\}$. Also, the cumulative distribution function of $T_k$ depends on $M$. We can show that
\begin{align*}
\Pr \left\{T_k \leq t \right\} = \sum_{m=1}^{\infty} \Pr \left\{T_k \leq t | M = m \right\} \Pr \left\{ M = m \right\}.
\end{align*}

If $X_k \leq r$, vehicle $k$ merges with vehicle $k-1$. The size of platoon $n-1$ becomes $M + 1$. If $X_k > r$, there is no cost of platooning. We can calculate the expectation of $T_k$ by
\begin{align*}
\mathrm{E}[T_k] = \sum_{m=1}^{\infty} \mathrm{E}[T_k | M=m] \Pr \left\{M=m \right\}.
\end{align*}

Since we assume $h_0 \approx 0$, we can simplify the expression of $T_k$ as
\begin{align*}
T_k = X_k + X_{k-1} + X_{k-2} + \cdots + X_{k-m+2}.
\end{align*}

We can use the technique in Proposition 2 to divide the expectation into independent parts. Then the expectation of time reduction is:
\begin{align*}
\mathrm{E}[T_k] & = \sum_{m=1}^{\infty} \Pr \left\{M=m \right\} \Big( \mathrm{E}[X_k | M=m] + \\
				& \quad \ \mathrm{E}[X_{k-1} | M=m] + \cdots + 
				\mathrm{E}[X_{k-m+2} | M=m] \Big) \\
				& = \sum_{m=1}^{\infty} \Pr \left\{M=m \right\} \Big( (m-1) \mathrm{E}[X_k | X_k \leq r] \Big).
\end{align*}

We apply the expectation equation to our Poisson arrival process, then the expectation equation is
\begin{align*}
\mathrm{E}[T_k] & = \sum_{m=1}^{\infty} e^{-\lambda r}(1-e^{-\lambda r})^{m-1} (m-1)(\frac{1}{\lambda} - \frac{r}{e^{\lambda r} - 1}) \\
				& = \frac{1}{\lambda} e^{\lambda r} - r - \frac{1}{\lambda}.
\end{align*}

\hfill $\blacksquare$ 
\subsection{Comparison with micro-simulation model}
To validate the results obtained from the stochastic model, we use the Simulation of Urban MObility (SUMO \cite{krajzewicz2002sumo}) to validate our results. This tool is designed for microscopic traffic simulation.
In accordance with our platooning policy characterized by $r$, the following vehicle merges with leading vehicle if inter-arrival time is less than given threshold. We use a PID controller to manage the movement of vehicles inside a platoon.

In our experiment, the highway length is $10.0 \ km$, and the simulation time step is $0.1 \ s$. In the validation of between platoon headway proposition, we use the headway between the first vehicle in following platoon and the last vehicle in leading platoon. For the calculation of vehicle time reduction, we subtract travel duration from travel time with free flow speed. In our setting, the free flow travel time is $414.0 \ s$. We have simulated $1000$ vehicles' arrival for each threshold $r$ in SUMO. The vehicle arrival in each simulation is a Poisson process.

In the SUMO simulation, a modified Krauß car-following model \cite{song2014research} is designed to manage vehicle behaviors. Compared with simulation only using Poisson process to generate vehicle departure, the car-following model would affect headway between platoons and time increment of platooning. In our experiment, the leading vehicle utilizes the car-following model to adjust its acceleration, which causes speed variations for vehicles.

\begin{figure}[ht]
  \centering
  \includegraphics[width=0.50\textwidth]{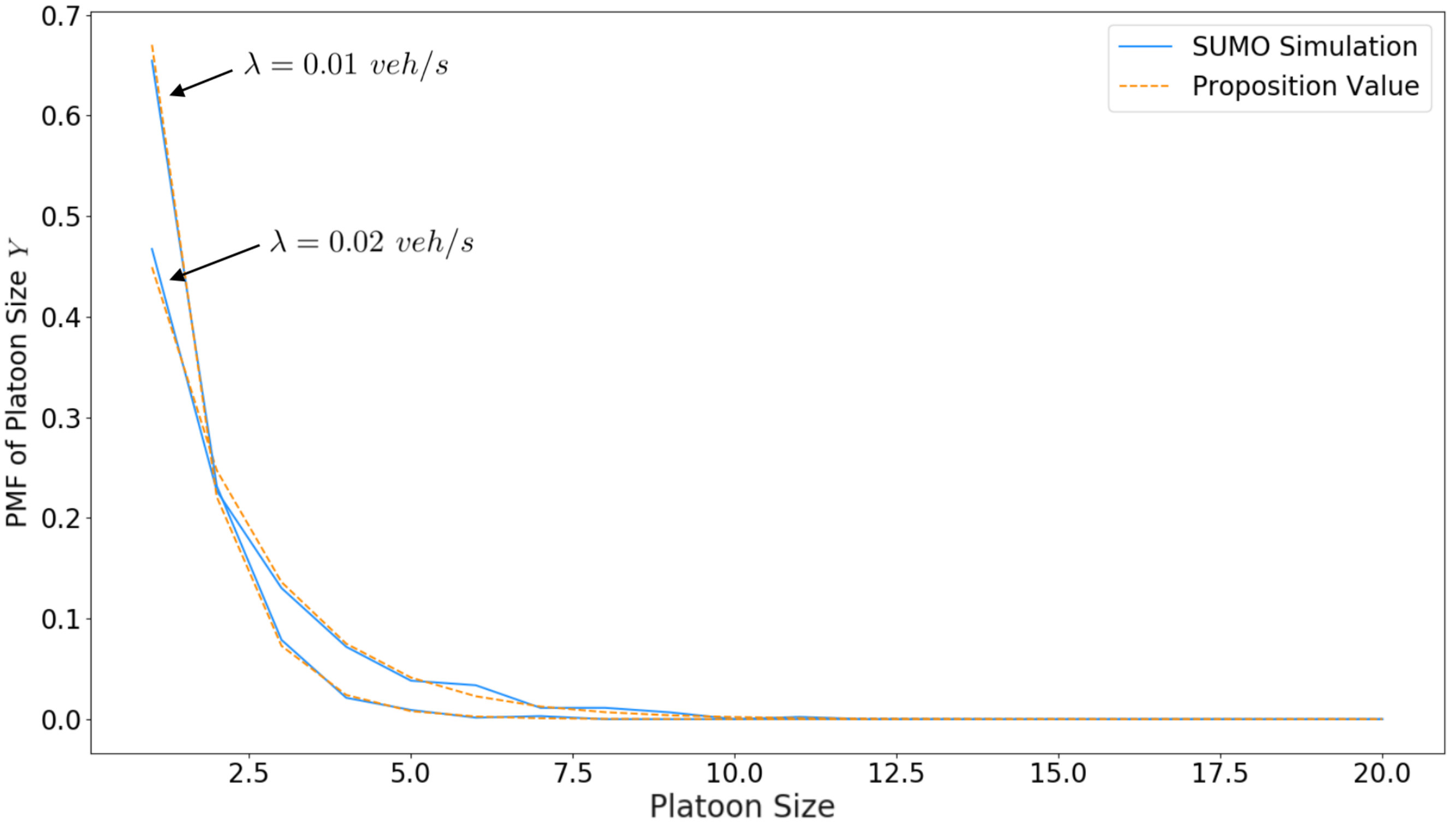}
  \includegraphics[width=0.50\textwidth]{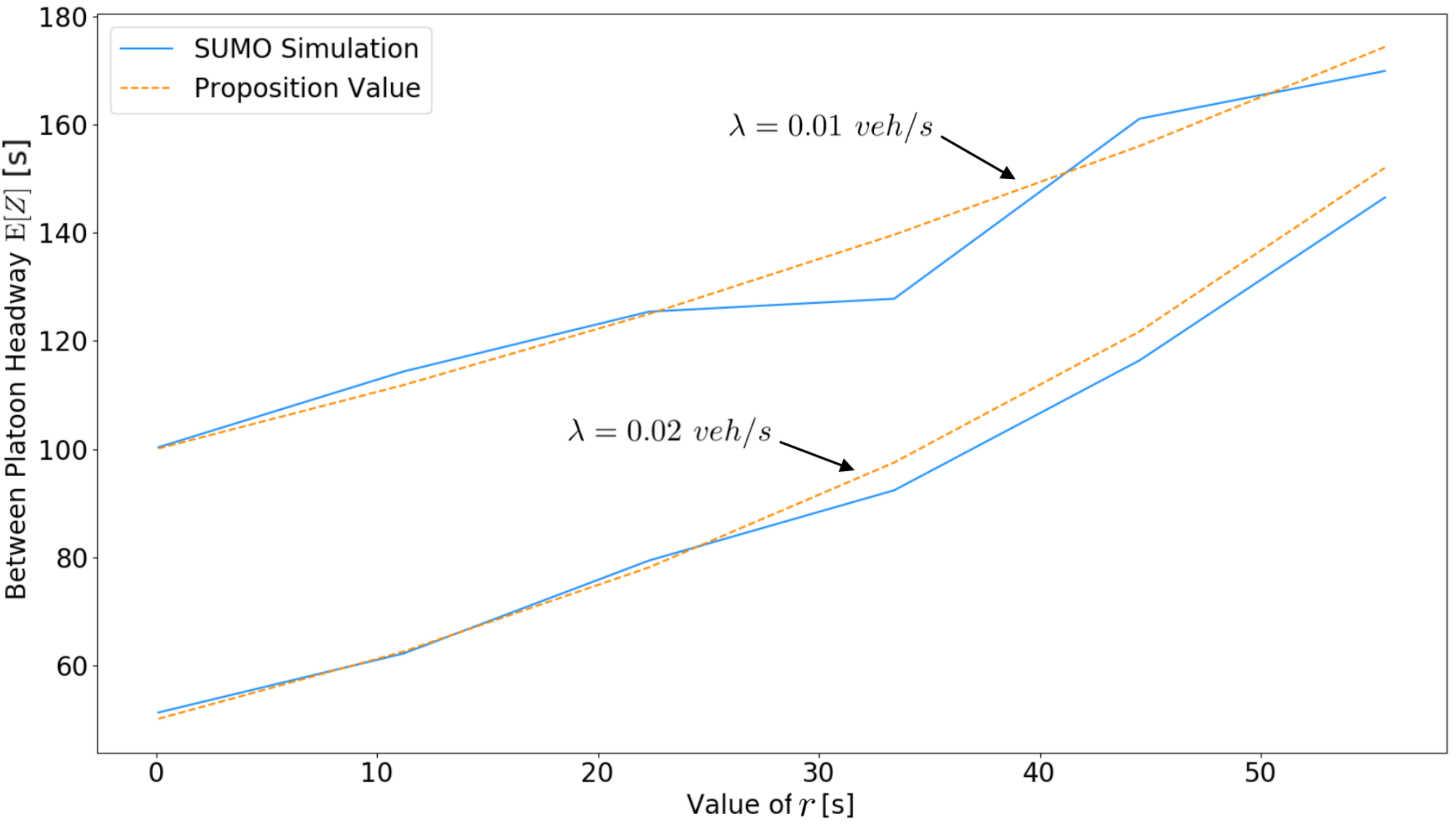}
  \includegraphics[width=0.50\textwidth]{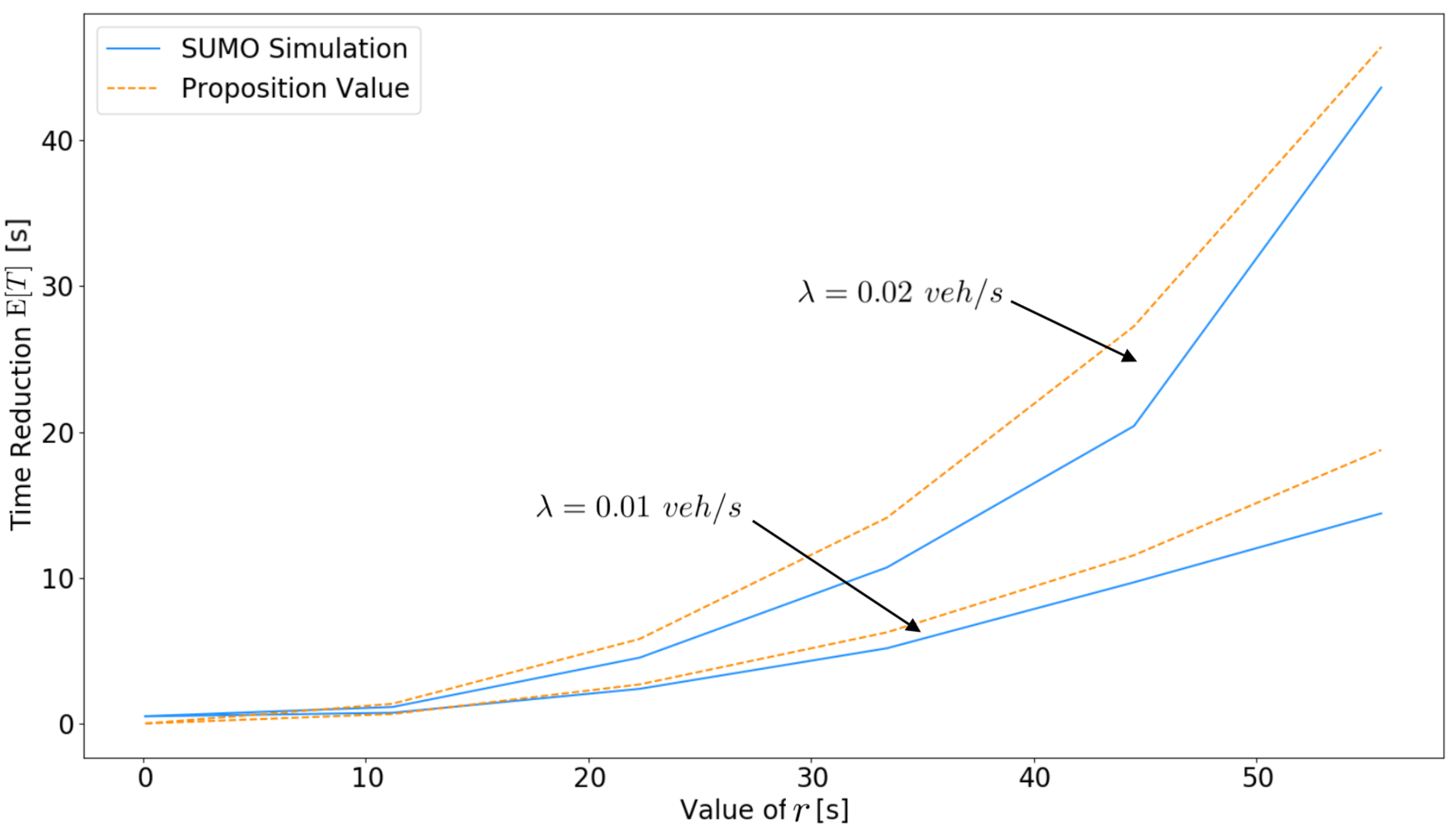}
  \caption{Compassion between SUMO simulation and Proposition values.}\label{fig:three_figures}
\end{figure}

Fig. \ref{fig:three_figures} shows the comparison between SUMO simulation and values in propositions under different arrival rates. In the figure of platoon size, the PMF drops dramatically with platoon size. In between platoon headway and time reduction figures, both values increase along with platooning threshold. Although our stochastic model does not consider the microscopic details in SUMO, its prediction is fairly consistent with SUMO results.

\section{Optimizing platooning threshold $r$}
\label{sec_optimize}

In this part, we analyze the effect of threshold $r$ on the total cost of platooning. The value of $r$ affects the feasibility of forming platoons. We analyze this in an economic perspective.

The benefits of platooning include two parts: reducing travel time and fuel consumption. We take the acceleration process for example. During the process of platooning, the time reduction increases while fuel consumption increases. After the merging process, vehicles in the platoon have less aerodynamic drag and thus experience less fuel consumption \cite{larson2015distributed}. We divide the total cost of platooning $TC_k$ into three parts: time reduction $T_k$, increased fuel $\Delta F_1$ during merging and fuel reduction $\Delta F_2$ after merging. The expected total cost can be expressed as:

\begin{align*}
\mathrm{E}[TC_k] = - w_1 \mathrm{E}[T_k] + w_2 \mathrm{E}[\Delta F_1] - w_2 \mathrm{E}[\Delta F_2].
\end{align*}
in which, $w_1$ is value of time (VOT) and $w_2$ is fuel price. We should notice that the total cost $TC_k$ is the incremental total cost instead of absolute value.

The derivation of $\mathrm{E}[T_k]$ has been shown in Proposition 3. We will show the calculation of fuel consumption in the following part.

In order to justify the effect of platooning in a qualitative way, we assume a scenario where the highway includes merging zone and platoon cruising section (Fig. \ref{fig:color_figure}). Merging zone is much smaller than normal cruising section. To simplify the modeling, we assume that the following vehicle always accelerates during the merging zone. Let the distance of merging be $D_1$ and the cruising distance after platooning be $D_2$. The average speed on highway is $v$.
We consider relative benefits of coordinated platooning and define the cost for vehicle $k$ as $F_k=\alpha D_k {v_k}^2$, which represents added fuel consumption from air drag based on distance and velocity \cite{larson2015distributed}. $\alpha$ is the coefficient of additional fuel consumption on distance and velocity. We express the increased fuel $\Delta F_1$ as:
\begin{align*}
\Delta F_1 = \alpha D_1 \left( \frac{D_1}{D_1/v-T_k} \right)^2 - \alpha D_1 v^2.
\end{align*}

In real scenarios, $T_k \ll D_1/v$. We can use the technique of linearization to approximate the expectation of $\Delta F_1$,
\begin{align*}
\mathrm{E}[\Delta F_1] & = \alpha \mathrm{E}[T_k] \frac{\partial \Delta F_1}{\partial T_k}\bigg|_{T_k = 0} \\
             & = 2 \alpha v^3 \mathrm{E}[T_k].
\end{align*}
Following vehicles in a platoon save a fraction $\eta \in (0,1)$ of fuel consumption when cruising alone \cite{larson2015distributed}. Let $\theta$ represent the fuel efficiency, we can show that:
\begin{align*}
\mathrm{E}[\Delta F_2] & = \eta \theta D_2 \Pr \left\{X_k \leq r \right\} = \eta \theta D_2 (1-e^{-\lambda r}).
\end{align*}
Summing up these three parts, total cost can be expressed as:
\begin{align*}
\mathrm{E}[TC_k] & = - w_1 \mathrm{E}[T_k] + 2 \alpha w_2 v^3 \mathrm{E}[T_k] - w_2 \mathrm{E}[\Delta F_2] \\
         & = (2 \alpha w_2 v^3 - w_1)(\frac{1}{\lambda} e^{\lambda r} - r - \frac{1}{\lambda}) \\
         & \quad \ - w_2 \eta \theta D_2 (1-e^{-\lambda r}).
\end{align*}

To minimize the total cost of platooning, we investigate the optimal value of $r$ by:
\begin{align*}
\frac{\partial \mathrm{E}[TC_k]}{\partial r} = \frac{(2 \alpha w_2 v^3 - w_1)\left( e^{2 \lambda r} - e^{\lambda r} \right) - D_2 w_2 \eta \theta \lambda }{e^{\lambda r}}.
\end{align*}

We analyze the results in two parts:
\begin{enumerate}
  \item When $2 \alpha w_2 v^3 - w_1 \leq 0$, we have $\frac{\partial \mathrm{E}[TC_k]}{\partial r} < 0$. $\mathrm{E}[TC_k]$ decreases as we have higher $r$;
  \item When $2 \alpha w_2 v^3 - w_1 > 0$, we let $\frac{\partial \mathrm{E}[TC_k]}{\partial r} = 0$. $\mathrm{E}[TC_k]$ gets the optimal value when $r=\frac{1}{\lambda} \ln \left( \frac{1}{2} + \frac{1}{2} \sqrt{\frac{4 D_2 w_2 \eta \theta \lambda}{2 \alpha w_2 v^3 - w_1} + 1} \right)$.
\end{enumerate}

We use nominal values in Table \ref{table:nominal_value} to justify the effect of $r$ on total cost.
\begin{table}[!ht]
  \caption{Nominal values in Poisson arrival process}\label{table:nominal_value}
  \begin{center}
    \begin{tabular}{|c|c|c|}
    \hline
    Parameter    &  Value                                   \\
    \hline
    $\eta$       &  $0.1$                                  \\
    $\lambda$    &  $0.02 \ veh/s$                           \\
    $\alpha$     &  $6.78\times10^{-7} \ L \cdot s^2 / m^3$      \\
    $\theta$     &  $41.0 \ L/100 km$                       \\
    $v$          &  $55 \ miles / hour$                     \\
    $w_1$        &  $25.8 \ \$ / hour$                        \\
    $w_2$        &  $0.868 \ \$ / L$   \\
    \hline
    \end{tabular}
  \end{center}
\end{table}

\begin{figure}[ht]
  \centering
  \includegraphics[width=0.45\textwidth]{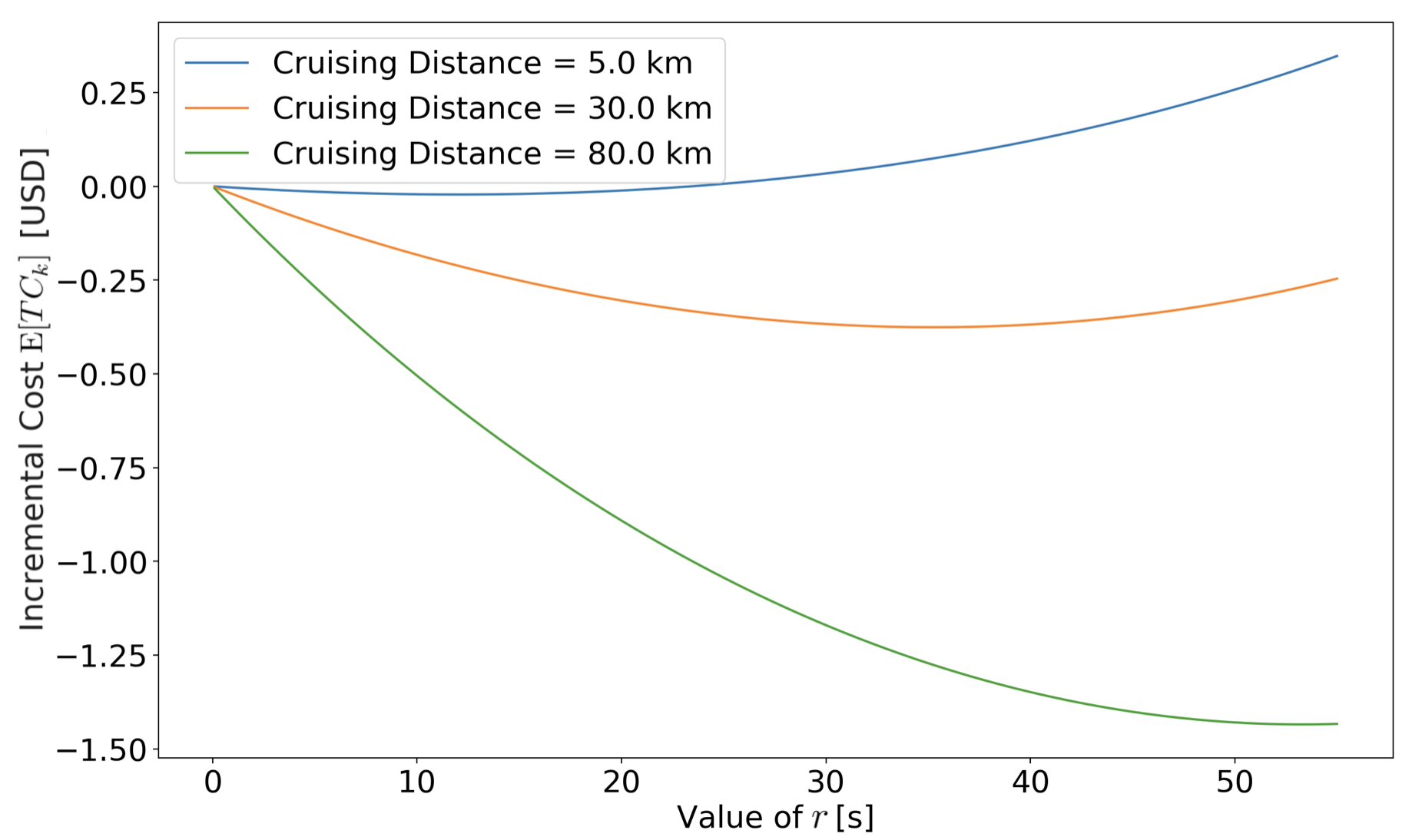}
  \caption{Effects of $r$ on cost of platooning for various driving distances over the cruising zone.}\label{fig:combine_distance}
\end{figure}

Intuitively, the total cost changes with cruising distance $D_2$. We chose three different cruising distances to present the effect of $r$ on total cost. The results are shown in Fig. \ref{fig:combine_distance}. From the figure we can see that when the cruising distance is small ($5.0 \ km$), increased fuel consumption $\Delta F_1$ plays a major part. Optimal threshold $r$ approximates $0$ in this case. Total cost almost increases with the value of $r$, which increases the accessibility of platooning. When forming more platoons, the reduction of time increases. However, the fuel cost during merging increases more rapidly. In this scenario, the value of total cost is positive, which means the benefits are negative. We tend to have lower $r$ to minimize the total cost.

When the cruising distance equals $30.0 \ km$, the proportion of fuel cost reduction after merging increases. The total cost first decreases and then increases at a turning point. There exists an optimal threshold of platooning as we have shown.

When the cruising distance increases to $80.0 \ km$, the fuel cost reduction after merging weighs more than two other parts. As the forming of platoons becomes more accessible, longer driving distances represent higher fuel savings. We tend to increase the threshold to minimize total cost. In this case, total cost is negative. We could have more benefits as the platoon operates on the highway.

\section{Concluding remarks}
\label{sec_conclude}

In this paper, we have developed a threshold-based method of platooning. We presented three platooning characteristics, including platoon size, expectation of headway between platoons and time reduction due to platooning. Then we analyzed optimal platooning threshold to minimize total incremental cost. The results show that the optimal platooning policy depends on the length of highway section. If the highway section is long enough, higher threshold and longer platoons are favorable; otherwise we should have shorter platoons.

This work can be extended in several directions. 
First, analysis of platoon characteristics can consider more general distributions of arrival headways (e.g. normal distribution).
Second, deceleration and the mix of acceleration/deceleration can be analyzed.
Third, more flexible threshold (e.g. platoon size-sensitive) can be incorporated. 

\bibliographystyle{IEEEtran}
\bibliography{bib_LJ}   
\end{document}